# Singularity softening and avoidance by the action of thermal radiation in a generalized entropic cosmology

***E. Elizalde*** [1]

[1] *Institute of Space Science, ICE/CSIC and IEEC,*
*Campus UAB, C/Can Magrans, s/n,*
*08193 Bellaterra, Barcelona, Spain*

***A. V. Yurov*** [2]

[2] *Institute of High Technology*
*Immanuel Kant Baltic Federal University (BFU)*
*Kaliningrad 236041, Russia*

***A. V. Timoshkin*** [3,4]

[3] *Tomsk State Pedagogical University (TSPU), 634041 Tomsk, Russia*

[4] *Lab. for Theor. Cosmology, International Centre of Gravity and Cosmos,*
*Tomsk State University of Control Systems and Radio electronics (TUSUR),*
*634050 Tomsk, Russia*

elizalde@ice.csic.es
aiurov@kantiana.ru
alex.timosh@rambler.ru

**Abstract.** Some relevant aspects of a new form of generalized entropic cosmology, recently introduced by Nojiri, Odintsov and Faraoni, are considered. The setup is a logarithmic equation of state for a viscous dark fluid coupled with dark matter, in the ordinary Friedmann-Lemaître-Robertson-Walker flat universe. The influence of thermal effects, caused by Hawking radiation, near the singularity, are carefully investigated. In particular, their role on the formation and specific type of the Big Rip expected to occur within a finite time. It is shown that a scenario arises, where a qualitative change towards the good direction, in the type of the singularity formed, does occur. On top of that, another very interesting scenario is obtained, where the singularity vanishes completely.

## I. Introduction

Recent results of astronomical observations confirm the fact that our universe is currently in accelerated expansion [1, 2]. One possible way to explain the origin and nature of such cosmic acceleration is the approach that makes use of a modified gravity theory based on the connection of gravity with thermodynamics. Within this framework, extended entropy is used instead of the usual Boltzmann-Gibbs entropy. As is well-known, the Boltzmann-Gibbs entropy is applicable to additive systems, while it is also known that gravitational systems do not have this property. As a consequence, it is necessary to generalize the Boltzmann-Gibbs entropy. As prospective generalizations, which are not extensive entropies, one can count the ones proposed by Tsallis [3, 4], Barrow [6], Renyi [7], Kaniadakis [8], Sharma-Mittal [9], and Loop Quantum Gravity [10]. A new generalized entropy function was recently introduced by Nojiri, Odintsov and Faraoni in [11]; and another model for cosmology, relying on a nonsingular entropy function has been proposed by Odintsov and Paul [12]. The nonsingular behavior of the entropy function allows to describe a corresponding bounce cosmology reasonably well.

The first and very visible consequence of the application of such thermodynamic approach in cosmology is a modification of the Friedmann equations, which is proven to disappear —giving back the usual Friedmann equations— when in the generalized entropy formula the limit yielding back the usual Boltzmann-Gibbs entropy is taken. In other words, the procedure is well under control.

If we consider the phantom era in the evolution of the universe, e.g., when the thermodynamic parameter in the equation of state (EoS) takes values that are less than minus one ($\omega < -1$), then the Universe evolution can take various ways. A characteristic feature of this era in the Universe development is the possible formation of singularities in the future. All these singularities, even if different in type, are classical in nature. In the dark energy era, the appearance of the singularity of the most destructive type, namely the Big Rip (BR), is possible at the end-time of the formation $t_{rip}$. A BR leads to the decay of gravitationally bound objects that are large on a cosmological scale.

There are other soft singularities, classified as Types II–IV. A general, comprehensive classification of finite-time future singularities was proposed in the seminal work by by Nojiri, Odintsov and Tsujikava [13] (for a review, see also [14-18]). According to this classification, there are four main different types of singularities.

At the occurrence of the singularities, the main cosmological parameters describing the Universe, as the scale factor $a(t)$, the effective (total) energy density $\rho_{eff}$, and the effective (total) pressure $p_{eff}$ in the limit $t \to t_s$, change as follows:

- Type I (BR): $a \to \infty$, $\rho_{eff} \to \infty$ and $|p_{eff}| \to \infty$. This class of singularities includes the case when $\rho_{eff}$ and $p_{eff}$ are finite at $t = t_s$. A Big Rip leads to the decay of gravitationally bound objects large on a cosmological scale.
- Type II (“sudden” singularity): $a \to a_s$, $\rho_{eff} \to \rho_s$ and $|p_{eff}| \to \infty$, where $a_s \neq 0$ and $\rho_s$ are constant. That is a pressure singularity.
- Type III: $a \to a_s$, $\rho_{eff} \to \infty$ and $|p_{eff}| \to \infty$. This type singularity is milder than Type I but stronger than Type II.
- Type IV: $a \to a_s$, $\rho_{eff} \to 0$ and $|p_{eff}| \to 0$, but the higher derivatives of the Hubble function $H$ diverge. This type also includes the case where $\rho_{eff}$ and/or $|p_{eff}|$ are finite for $t = t_s$.

The effective energy density $\rho_{eff}$ and the effective pressure $p_{eff}$ are given by the expressions

$$\rho_{eff} = \frac{3}{k^2} H^2, \ p_{eff} = -\frac{1}{k^2}\left(2\dot{H} + 3H^2\right), \tag{1}$$

where $k^2 = 8\pi G$, being $G$ the Newton constant and $H = \frac{\dot{a}}{a}$ the Hubble parameter.

Note however that the singularity is not the only possible ending of the evolution of the universe in the phantom phase. For example, cosmological models exist with either a Little Rip or a Pseudo Rip, and they are non-singular. In these cases, in the far future, depending on the asymptotic behavior of the Hubble parameter $H$, the dark-energy density increases monotonically or either remains constant [19-22].

In the following, we will investigate the influence of thermal effects caused by the Hawking radiation on the formation of a singularity of the BR type in the cosmological model here considered, that is, in the framework of the new generalized entropy in Ref. [11], taking into due account the bulk viscosity of the dark fluid and its interaction with dark matter.

We shall start by discussing some physical aspects of the Hawking radiation, considering the behavior of the universe in the vicinity of the singularity point. It is known that an increase of the Hubble parameter near the singularity leads to an increase of the universe temperature. As the result of such high temperature, thermal radiation appears. Thermal radiation is here caused by the Hawking radiation generated on the visible horizon of the Friedmann-Lemaître-Robertson-Walker (FLRW) universe [23-26]. The spectrum of thermal radiation is formed at high temperatures in the late Universe, immediately before the emergence of the rip. Corrections associated with thermal radiation, and the assumption about viscosity of the cosmic fluid, make a classical description of the universe near the rip more accurate. The property of the viscosity of the cosmic fluid also affects the behavior of the universe near the singularity. Viscosity must be duly considered, when describing not only singularities of the BR type [27], but also other possible types of singularities: II, III and IV [28, 29]. It allows to predict the future state of the universe.

We will use an entropic cosmological model, in particular, a description based on a new generalized entropy function. This approach to a modified theory of gravity is specifically based on the connection between gravity and thermodynamics. According to entropic cosmology, the entropy function generates the energy density and pressure in the Friedmann equations. As a consequence, the Friedmann equations are derived from the fundamental laws of thermodynamics, a well-known result. According to this thermodynamical approach, at the late stage of the evolution of the universe, in the case when the cosmological constant is equal to zero, the dark energy density is entirely generated from the entropy energy density. And as it turns out, the theoretical values of the parameters in the EoS do agree with the results of the most accurate astronomical observations from the Planck satellite. Notwithstanding that, the presence of a cosmological constant in the entropic dark energy model makes it more viable. As a consequence, generalized entropy cosmology also includes the cosmological constant case, as a special one.

As follows from the results of the astronomical observations, the standard cold dark matter model provides very efficient results at very large (cosmological) scales, but its application might be problematic at galactic scales. This problem follows from the assumption that dark matter has no pressure. However, the description of the late-time Universe at small scales can be obtained within the framework of the logarithmically-corrected equation of state for the matter sector, within the Debye approximation [30]. Following such model, the fluid pressure can be represented in the form of an empirical formula for the pressure of crystalline solids deformed under isotropic stress [31]. In order to describe the accelerated expansion of the universe, it is necessary that the pressure of the fluid specified in terms of the EoS is negative. The pressure of the dark fluid, presented in terms of the logarithmically corrected EoS, becomes dominant provided the volume of the universe exceeds a certain value. This case corresponds to the approaches of the logo tropic dark energy scenario LDE [32, 33].

Before concluding this introduction, acknowledgment should be given to similar researches carried out on this topic. Theoretical applications of the theory of a one-component viscous fluid with a logarithmically corrected EoS were investigated in [34]. And the description of a two-component fluid, containing an interaction term with dark matter, in the same kind of model, was considered in [35].

In the present article, we will apply a thermodynamic approach to the study of the evolution of the late Universe, on the basis of new generalized entropy proposed in Ref. [11]. The description will be given on the basis of a logarithmically adjusted EoS for a dark fluid, by taking into account bulk viscosity and a generalized Friedmann equation.

The main aim of this work will be to study the effect of thermal radiation very shortly before the singularity is formed, by considering the influence of the properties of the dark fluid's viscosity, and its coupling with dark matter, on eventual qualitative changes that are expected to take place in the forming singularity of the BR type.

## II. Viscous dark fluid with modified log-corrected power-law equation of state

In this section we shall start with the theoretical formulas corresponding to the logarithmic EoS in the logarithmically corrected power-law theory [30]. For the pressure, we have

$$p = A\left(\frac{\rho}{\rho_*}\right)^{-l} \ln\left(\frac{\rho}{\rho_*}\right). \tag{2}$$

Which is an empirical formula for the pressure of deformed isotropic crystalline solids. In order for the universe to expand with acceleration, it is necessary that the dark fluid pressure described by the EoS be negative. The negative pressure in the log-corrected power-law model becomes dominant, provided the volume of the universe exceeds a certain value. This scenario for the evolution of the Universe is called logarithmically corrected power-law dark energy (LDE). As a consequence, Eq. (2) can describe dark fluid pressure in a cosmological model of an expanding universe. The modified logarithmic EoS in $F(R)$ gravity was first used in Ref. [31].

In Eq. (2), $\rho_*$ is a reference density, to be identified with the Planck density [32]: $\rho_p = c^5/\hbar G^2 \approx 5.16\times 10^{99}\, gr/m^3$. The parameter $A>0$ represents the logotropic temperature, and $l = -\frac{1}{6} - \gamma_G$, where $\gamma_G$ is the non-dimensional Gruneisen parameter. The parameter $A$ is determined by the relation $A = B\rho_\Lambda c^2$ [32]. The dimensionless logotropic temperature is approximately given by $B \approx 1/ln(\rho_P/\rho_\Lambda) \approx 1/[123 ln(10)]$, where $\rho_P$ and $\rho_\Lambda$ are the Planck density and the cosmological density, respectively. Consequently, as a new interpretation, the number $123 \simeq log(\rho_P/\rho_\Lambda)$ may look like the inverse logotropic temperature. The special case $l = 0$ in Eq. (2) does correspond to the logotropic cosmological model [30].

Let us discuss the condition under which the pressure in Eq. (2) becomes negative. For this purpose, we express the mass density in terms of the volume, using the inverse dependence $V \sim \frac{1}{p}$. Then, in the Debye approximation, one can write the Anton-Schmidt's pressure (2) in the form [36]:

$$p(V) = -\beta\left(\frac{V}{V_0}\right)^{-\frac{1}{6}-\gamma_G} \ln\left(\frac{V}{V_0}\right). \tag{3}$$

Here $V$ is the volume of the crystal and $V_0$ is a given, reference volume representing a barrier, which the Universe must overcome to ensure that accelerated cosmic expansion takes place. The parameter $\beta$ is the bulk modulus at value $V_0$.

Consider the equation $\rho_{eff} + 3p_{eff} = 0$. It is of interest, in this case, to analize the transition of the universe, from deceleration to acceleration, depending on the values of the Gruneisen parameter. Namely, if $\gamma_G \le \frac{5}{6}$, then in the process of expansion there is always a transition from deceleration to acceleration; when $\gamma_G > \frac{5}{6}$, the situation is more interesting and depends on the sign of the logarithm of $K = \frac{3\beta V_0}{\sigma\left(\gamma_G - \frac{5}{6}\right)}$, where the parameter $\sigma$ is always positive. According to this, three different situations appear:

- when $\ln K < 1$, the acceleration does not change its sign;
- when $\ln K = 1$, the situation is qualitatively similar to the case $\gamma_G \le \frac{5}{6}$, and there is a single change of sign;
- when $\ln K > 1$, during the expansion process, the second derivative of the scale factor $\ddot{a}$ changes sign twice: there is a transition from deceleration to acceleration, followed by deceleration.

For an ideal gas, the Gruneisen parameter is $\gamma_G = \frac{2}{3} < \frac{5}{6}$; that is, the process of expansion does take place with one transition from deceleration to acceleration. In the following, we will assume that in the vicinity of the transition points, the energy density can be neglected in comparison with the effective pressure.

We now consider Eq. (3). Again, three different cases appear:

1. The pre-barrier era, when $V < V_0$. In this case the pressure is positive and the universe is decelerating.
2. The time of equivalence between $V$ and $V_0$. In this case there exists a transition from deceleration to acceleration of the universe, occurring exactly when $V = V_0$.
3. The post-barrier era, e.g., the time after passing the $V_0$ barrier, when $V > V_0$. In this case the pressure is negative and the fluid starts to accelerate the universe.

In these regimes, the log-corrected power-law dark fluid model is equivalent to an LDE model. This circumstance allows us to describe the evolution of the late universe by using the log-corrected power-law EoS.

Now, we will consider the dynamical evolution of the late universe using a viscous dark fluid, coupled with dark matter. With this aim, the EoS Eq. (2) will be modified, by including a term with bulk viscosity $\zeta(H,t)$, given by [37]:

$$\zeta(H,t) = \xi_1(t)(3H)^n, \tag{4}$$

which depends on the Hubble function, $H$, and on time, $t$. The function $\xi_1(t)$ is an arbitrary function that depends on time. Then, Eq. (2) takes the form:

$$p = A\left(\frac{\rho}{\rho_*}\right)^{-l} \ln\left(\frac{\rho}{\rho_*}\right) - 3H\zeta(H,t). \tag{5}$$

This equation is the analog of the EoS for a dark viscous fluid and it is a particular case of the generalized EoS. A review of the dark fluid cosmology can be found in [38].

## III. Properties of generalized entropy

The generalized entropy function $S_g$ to be considered here has the form [11]

$$S_g(\alpha_+,\alpha_-,\beta,\gamma) = \frac{1}{\gamma}\left[\left(1+\frac{\alpha_+}{\beta}S\right)^{\beta} - \left(1+\frac{\alpha_-}{\beta}S\right)^{-\beta}\right], \tag{6}$$

where $\alpha_+,\alpha_-,\beta,\gamma$ are positive parameters and $S$ is the Bekenstein-Hawking entropy. As is well-known, this entropy describes the radiation from a black hole [23, 39]

$$S = \frac{C}{4G}, \tag{7}$$

where $C = 4\pi r_h^2$ is the horizon area, $r_h$ is the horizon radius and $G$ is Newton's gravitational constant. Other entropy functions, as those of Tsallis, Barrow, Renyi, Kaniadakis and Sharma-Mittal, reduce in some limit to the Bekenstein-Hawking entropy, and are monotonically increasing in the variable S.

Moreover, the generalized entropy function $S_g$ reduces to those of Tsallis, Barrow, Renyi, Kaniadakis and Sharma-Mittal, with appropriate choices of the parameters.

The main properties of the function $S_g$ are as follows:

1. The entropy $S_g(\alpha_+,\alpha_-,\beta,\gamma)$ satisfies the generalized third law of thermodynamics $S_g \to 0$, when $S \to 0$.
2. The entropy $S_g(\alpha_+,\alpha_-,\beta,\gamma)$ is a monotonically increasing function in the variable $S$, because both terms, $\left(1+\frac{\alpha_+}{\beta}S\right)^{\beta}$ and $\left(1+\frac{\alpha_-}{\beta}S\right)^{-\beta}$ in the expression of $S_g$ increase with the variable $S$.

3. The entropy $S_g\left(\alpha_+,\alpha_-,\beta,\gamma\right)$ reduces to the Bekenstein-Hawking entropy for a certain limit of the parameters.

In particular, when the parameters in the generalized entropy function $S_g$ take the values: $\alpha_+\to\infty,\alpha_-=0$, $\gamma=\left(\frac{\alpha_+}{\beta}\right)^{\beta}$ and $\beta=1$, then the generalized entropy function approaches the Bekenstein-Hawking entropy: $S_g\to S$.

As shown in [11] in detail, the minimum number of parameters in the generalized entropy function to describe all entropies listed above is equal to four.

## IV. The effect of thermal radiation on the formation of singularities in entropic cosmology, for a viscous dark fluid coupled with dark matter

In this section we consider a spatially-flat (FLRW) metric and apply the thermodynamic approach to describe the evolution of the universe based on the generalized entropy function $S_g$

$$ds^2=-dt^2+a^2\left(t\right)\sum_{i=1,2,3}\left(dx^i\right)^2. \tag{8}$$

Here $a\left(t\right)$ is a scale factor. We will study the phantom phase of the late universe, when the thermodynamic parameter $\omega$ in the (EoS) is less than -1. In this phase, future finite time singularities can appear. We will be interested in the influence of thermal radiation, which occurs near the singularity, on a qualitative change of the Big Rip singularity. A cosmological singularity of the (BR) type is most destructive. It is associated with the phantom era of the evolution of the universe. In this case, the universe expands very quickly. However, any expanded object will be destroyed millions of years before the rip time. All types of singularities are purely classical singularities in nature.

As is known, from a physical point of view, an increase of the Hubble function leads to an increase of the temperature. As a result, at the high temperatures near the singularity, thermal radiation appears. The nature of this thermal radiation is associated with Hawking's radiation, which is generated on the apparent horizon of the FLRW universe. Hawking's radiation appears in black holes and is associated with the existence of the visible horizon of the black hole, as well as the visible horizon of cosmic events in de Sitter space. Hawking's thermal spectrum radiation shows up in the late universe, at high temperatures, shortly before its Rip time. Accounting for thermal radiation will allow a qualitative change in the classical description of the singularity and creates a more realistic picture of the future universe.

From statistical physics, it follows that the thermal radiation energy density is proportional to the fourth power of the absolute temperature. Thus, near the future singularity, where the Hubble parameter increases without limit, one can phenomenological represent the dark energy density as

$$\rho_{rad}=\lambda H^4, \tag{9}$$

where $\lambda$ is a positive constant.

Note, that in general case , the temperature of the horizon in cosmology is given by Hayward-Kodama temperature which contains both the Hubble parameter and its derivative. In our article we consider a late universe in the phantom stage of evolution, when the thermodynamic parameter in the equation of state for a dark fluid $\omega<-1$, which often leads the evolution of the universe to a finite-time future singularity like the Big Rip. If the universe continues to expand with acceleration (specified by the cosmological constant), the universe will approach de Sitter space time, where there will be a cosmological horizon that radiates in the same way as the event horizon of a black hole generates Hawking radiation. Thus, the asymptotic case described by equation (9) occurs.

The modified Friedmann equation takes the form [40]

$$\frac{3}{k^2}H^2=\rho_{eff}+\rho_g+\lambda H^4 . \tag{10}$$

where $H=\frac{\dot{a}(t)}{a(t)}$ is the Hubble function, $k^2=8\pi G$ is Einstein's gravitational constant. Here $\rho_{eff}$ is the effective energy density, $\rho_g$ denotes the energy density corresponding to the entropy function $S_g$

$$\rho_{eff}=\rho+\rho_m , \tag{11}$$

where $\rho$ is the dark energy density, $\rho_m$ is dark matter density. A dot denotes derivative with respect to the cosmic time $t$.

The energy density $\rho_{eff}$ and the pressure $p_{eff}$ can be calculated by using the following expressions:

$$\rho_{eff}=\frac{3}{k^2}H^2,\ p_{eff}=-\frac{1}{k^2}\left(2\dot{H}+3H^2\right). \tag{12}$$

Note, that the effective energy and pressure density may include a contribution from modified gravity.

The estimation of the Hubble parameter in the late time universe is of the order of $10^{-40}GeV$, and then the condition $GH^2 \ll 1$ is satisfied. The parameter $\gamma$ in (6) in the late time universe becomes constant $\gamma=\gamma_0$, then in this approximation for energy density $\rho_g$ looks as [40]:

$$\rho_g=\frac{3H^2}{k^2}\left[1-\frac{\alpha_+}{\gamma_0(2-\beta)}\left(\frac{GH^2\beta}{\pi\alpha_+}\right)^{1-\beta}\right]. \tag{13}$$

Then, we consider the simple case $\beta=1$, and obtain

$$\rho_g=\frac{3H^2}{k^2}\left(1-\frac{\alpha_+}{\gamma_0}\right). \tag{14}$$

We analyze the equation (10). To do this, taking into account Eq. (13), we solve it with respect to the square of the Hubble function

$$H^2=\frac{1}{2\lambda}\left[\frac{3\alpha_+}{\gamma_0 k^2}\pm\sqrt{\frac{9\alpha_+^2}{\gamma_0^2k^2}-4\lambda\rho_{eff}}\right]. \tag{15}$$

In the following, we will consider the physical consequences of the influence of thermal radiation on the formation of a singularity of the (BR) type. We will study the dynamical evolution of a late-time universe, applying a modified log-corrected power-law equation of state (5) in the presence of a bulk viscosity. To describe the accelerated expansion of the Universe in more detail, we will apply the model of a two-component coupled fluid. As the second fluid component, we will consider pressure-less dark matter weakly interacting with a log-corrected power-law dark fluid. We will apply the cosmological models from [35, 46].

Below, we consider various forms for the bulk viscosity.

***4.1 Constant viscosity***

Consider a log-corrected power-law coupled fluid in presence of a constant bulk viscosity $\zeta(H,t)=\zeta_0$ , in the approximation $\rho>\frac{\rho_*}{2}$, for the case when $l=-1$. Then, we can write the Hubble function in the form [46]

$$H=\frac{1}{a_1 t+C_1}+\frac{a_2}{2a_1}=\frac{1}{\frac{1}{4}\left(1-\frac{\lambda r}{1+r}\right)t+C_1}+\frac{\gamma_0\rho_*}{12\zeta_0}, \tag{16}$$

where $C_1$ is an arbitrary constant. We have introduced the notation $a_1=\frac{1}{4}\left(1-\frac{\lambda r}{1+r}\right)$, $a_2=\frac{3\zeta_0}{2\eta}$, where $\eta=\frac{\alpha_+}{\gamma_0(1+r)k^2}$. If $t$ tends to a finite value $t\to t_s=-\frac{C_1}{\frac{1}{4}\left(1-\frac{\lambda r}{1+r}\right)}$, then $H\to\infty$, and a cosmological singularity of the Big Rip type appears. In the limit $t\to\infty$ the Hubble function tends to the cosmological constant $H\to\frac{\gamma_0\rho_*}{12\zeta_0}$.

Now, we shall investigate the influence of thermal radiation on the qualitative change in the type of singularity and on the time of its appearance. Let's write the scale factor as

$$a(t)=a_0\left(a_1 t+C_1\right)^{\frac{1}{a_1}}e^{\frac{a_2}{2a_1}t}, \tag{17}$$

where $a_0$ is an integration constant. We consider the case $\lambda=\frac{1}{r}(1+r)\left(1+\frac{1}{3}r\right)$, then $a_1=1$. The effective energy density in terms of the scale factor will be expressed as follows:

$$\rho_{eff}=\frac{3}{k^2}\left(\frac{1}{t+C_1}+\frac{1}{2}a_2\right)^2. \tag{18}$$

Return now to Eq. (15). Since the function $H^2$ is a real, we must have

$$\frac{9\alpha_+^2}{\gamma_0^2k^4}-4\lambda\rho_{eff}\geq 0. \tag{19}$$

And, in this model, we obtain

$$\left(\frac{3\alpha_+}{\gamma_0k^2}\right)^2-\left(\frac{2\sqrt{3}\lambda}{k}\right)^2\left(\frac{1}{t+C_1}+\frac{1}{2}a_2\right)^2\geq 0. \tag{20}$$

Consider now the case $C_1=0$ and then from inequality (20), a restriction on the scale factor appears

$$a(t)\leq\frac{2\sqrt{\lambda}\gamma_0k}{\sqrt{3}\alpha_+}e^{\frac{3\zeta_0}{4\eta}t}. \tag{21}$$

Taking into account the thermal radiation $\rho_{rad}$, we obtain from Eq. (15) that there exists an upper limit $a_{\max}$ for the scale factor

$$a_{\max} = a(t_{\max}) = \frac{2}{\sqrt{3}} a_0 \frac{\sqrt{\lambda}\gamma_0 k}{\alpha_+} e^{\frac{1}{2}\sqrt{3\lambda}\frac{\zeta_0\gamma_0 k}{\alpha_+\eta}} , \tag{22}$$

which corresponds to the time $t_{\max}$,

$$t_{\max} = \frac{2\sqrt{\lambda}\gamma_0 k}{\sqrt{3}\alpha_+} \tag{23}$$

The time $t_{\max}$ corresponds to a new singularity. The difference between the singularity times $t_{\max}$ and $t_s = 0$ is positive, which means that $t_{\max}$ is larger than $t_s$. In the limit $t \to t_{\max}$, the scale factor tends to $a \to a_{\max}$.

Using Eqs. (12) one can calculate the effective energy density $\rho_{eff}$ and the effective pressure $p_{eff}$. Then, in the limit $t \to t_{\max}$, we obtain the expressions for the effective energy density and effective pressure, as

$$\rho_{\max} = \rho_{eff}(t_{\max}) = \left\{ \frac{3}{2} \left[ \frac{\alpha_+}{\sqrt{\lambda}\gamma_0 k^2} + \frac{\sqrt{3}}{2} \frac{(1+r)k\gamma_0\zeta_0}{\alpha_+} \right] \right\}^2 \tag{24}$$

and

$$p_{\max} = p_{eff}(t_{\max}) = \left| A\left( \frac{3H_{\max}^2}{\rho_* k^2} - 1 \right) - 9\tau H_{\max}^2 \right| = \left| \left( \frac{3A}{\rho_* k^2} - 9\tau \right) \left[ \frac{2\sqrt{3}\alpha_+}{\sqrt{\lambda}\gamma_0 k \left( 1 - \frac{\lambda r}{1+r} \right)} - \frac{\gamma_0\rho_*}{12\zeta_0} \right]^2 - A \right| . \tag{25}$$

The results obtained are as follows: because all, the scale factor, effective energy density and pressure, at the new singularity time $t_{\max}$ are finite, while higher derivatives of the Hubble function $H$ do not diverge, no cosmological finite-time future singularity in this model is formed.

### *4.2 Viscosity is proportional to the Hubble parameter*

In this section we shall consider the log-corrected power-law coupled fluid model with bulk viscosity to be proportional to the Hubble function

$$\zeta(H, t) = 3\tau H , \tag{26}$$

where the parameter $\tau$ is positive. As before, we assume that the density of the log-corrected power-law fluid is higher than the Planck density $\left( \rho > \frac{1}{2}\rho_* \right)$, and study the case $l = -1$.

We write the Hubble function under the form [46]

$$H = \sqrt{\frac{A}{\tilde{b}}} \frac{C_2 e^{2\frac{\sqrt{A\tilde{b}}}{\tilde{a}}t} + 1}{C_2 e^{2\frac{\sqrt{A\tilde{b}}}{\tilde{a}}t} - 1} , \tag{27}$$

where $C_2$ is an arbitrary constant and we have introduced the notation:

$$\tilde{a} = \frac{2\alpha_+}{\gamma_0 k^2} \left( \frac{1}{1+r} + \delta \right), \ \tilde{b} = \frac{3\alpha_+}{\gamma_0 (1+r) k^2} \left[ \frac{A}{\rho_*} + \lambda(1+r) + 1 \right] - 9\tau . \tag{28}$$

Let us now take $C_2 = 1$, then in the far future, in the limit $t \to \infty$, the Hubble function tends to the cosmological constant, $H \to \sqrt{\frac{A}{\tilde{b}}}$.

If $t$ tends to the finite value $t \to t_s = 0$, then the Hubble function increases indefinitely, $H \to \infty$, and a cosmological singularity of the BR type appears.

Consider again the behavior of the late-time universe near the singularity, taking into account the effect of thermal radiation, the scale factor being

$$a(t) = a_0 \left( sh \frac{\sqrt{A\tilde{b}}}{\tilde{a}} t \right)^{\frac{\tilde{a}}{\tilde{b}}}. \tag{29}$$

This can be written as $\tilde{a} = 2\tilde{b}$, and then $A = \rho_* \left[ (1+r)\left( \frac{\delta}{3} + \frac{3\tau\gamma_0 k^2}{\alpha_+} - \lambda \right) - \frac{2}{3} \right]$. In this case, the Hubble function and the scale factor take the form

$$H(t) = \sqrt{\frac{A}{\tilde{b}}} cth\left( \frac{1}{2}\sqrt{\frac{A}{\tilde{b}}} t \right), \; a(t) = a_0 \left( sh \frac{1}{2}\sqrt{\frac{A}{\tilde{b}}} t \right)^2. \tag{30}$$

Expressing the effective energy density in terms of the scale factor

$$\rho_{eff} = \frac{3A}{\tilde{b}k^2}\left[ 1 + \frac{a(t)}{a_0} \right], \tag{31}$$

from the inequality (19), it follows that:

$$a(t) < a_{\max} = a_0 \frac{\tilde{b}}{\lambda A}\left( \frac{\sqrt{3}\alpha_+}{\gamma_0 k} \right)^2. \tag{32}$$

The maximum number of the scale factor $a_{\max}$ corresponds to time

$$t_{\max} = 2\sqrt{\frac{\tilde{b}}{A}} arsh\left( \frac{1}{2}\sqrt{\frac{3\tilde{b}}{\lambda A}} \frac{\alpha_+}{\gamma_0 k} \right). \tag{33}$$

The time $t_{\max}$ corresponds to the occurrence of a new singularity. We see that the time of possible occurrence of singularities has again changed. The difference between singularity times $t_{\max}$ and $t_s = 0$ is positive. That is, in this model the singularity appears later.

Calculate now, in the limit $t \to t_{\max}$, the effective energy density and effective pressure

$$\rho_{\max} = \rho_{eff}(t_{\max}) = \frac{3A}{\tilde{b}k^2}\left[ 1 + \frac{\lambda A}{\tilde{b}}\left( \frac{2\gamma_0 k}{\sqrt{3}\alpha_+} \right)^2 \right], \tag{34}$$

$$p_{\max} = p_{eff}(t_{\max}) = A\left\{ \frac{1}{\tilde{b}}\left( \frac{3A}{\rho_* k^2} - 9\tau \right)\left[ 1 + \frac{\lambda A}{\tilde{b}}\left( \frac{2\gamma_0 k}{\sqrt{3}\alpha_+} \right)^2 \right] - 1 \right\}. \tag{35}$$

As in the previous model, the values of the scale factor, energy density and effective pressure turn out to be finite, and higher derivatives of the Hubble function do not diverge. According to the classification of singularities, again, no cosmological finite-time singularity is here formed.

However, if the parameter $\tilde{b}$ tends to zero, then the scale factor is finite, but the effective energy and pressure diverge. This case corresponds to a type III singularity, which is milder than a Type I, and stronger than a Type II. A transition from singularity types, from Type I to Type III, is here possible, due to the influence of thermal radiation. An interesting qualitative change in the singularity type occurs. However, the time of occurrence of the singularity did not change, in this case.

### *4.3 Viscosity with linear time dependence*

Consider now the viscosity term (4) in the case $n=1$ and choose the function $\xi_1(t)$ to vary linearly with time, namely

$$\xi_1(t)=\mu(ct+d), \tag{36}$$

where $\mu, c$ are dimensional parameters.

The solution of the gravitational equation of motion in the case $l=-1$, under the condition $\rho>\frac{\rho_*}{2}$, is [35]

$$H(t)=\frac{4\tilde{c}}{3\left(\tilde{c}t+\tilde{d}+\frac{A}{\rho_*}\right)+C_3}, \tag{37}$$

were $\tilde{c}=-3c(1+r)$, $\tilde{d}=1-c(1+r)k+\frac{3\alpha r}{(1+r)}$ and $C_3$ is as an integration constant.

Without loss of generality, consider the case $C_3=0$. The Hubble function diverges at a finite time $t_s=-\frac{1}{\tilde{c}}\left(\tilde{d}+\frac{A}{\rho_*}\right)$, and a Big Rip type singularity appears.

The scale factor is

$$a(t)=a_0\exp\left[-\frac{4}{3}\left(\tilde{c}t+\tilde{d}+\frac{A}{\rho_*}\right)^{-1}\right]. \tag{38}$$

And, representing the energy density through the scale factor

$$\rho_{eff}=\frac{1}{3}\left[\frac{\tilde{c}}{k}\ln^2\frac{a(t)}{a_0}\right]^2 \tag{39}$$

from the inequality (19), one gets a restriction on the scale factor

$$\frac{3\alpha_+}{\gamma_0 k}-2\sqrt{3\lambda}c(1+r)\ln^2\frac{a(t)}{a_0}\geq 0, \tag{40}$$

which, in this model, yields

$$a(t)\leq a_0 e^{\sqrt{\frac{\sqrt{3}\alpha_+}{2\sqrt{\lambda}\gamma_0 kc(1+r)}}}=a_{\max}. \tag{41}$$

The maximum number of the scale factor $a_{\max}$ corresponds to time

$$t_{\max}=\frac{1}{3c(1+r)}\left[\frac{4}{3}\sqrt{\frac{2\sqrt{\lambda}\gamma_0 kc(1+r)}{\sqrt{3}\alpha_+}}+\tilde{d}+\frac{A}{\rho_*}\right]. \tag{42}$$

Because $t_{\max}>t_s$, then the singularity may occur later.

In the limit $t\to t_{\max}$ the effective energy density and the effective pressure are, respectively,

$$\rho_{\max} = \rho\left(t_{\max}\right) = \frac{27\sqrt{3}}{2}\frac{\alpha_{+}c\left(1+r\right)}{\gamma_0\sqrt{\lambda}k^3}, \tag{43}$$

$$p_{\max} = \frac{27\sqrt{3}}{2}\frac{\alpha_{+}c\left(1+r\right)}{\gamma_0 k\sqrt{\lambda}}\left[\frac{A}{\rho_{*}k^2} - 9\mu\left(ct_{\max}+d\right)\right] - A. \tag{44}$$

In this case, the maximum numbers of the scale factor, of the effective dark energy density and effective pressure at the new singularity time $t_{\max}$ are finite, while higher derivatives of the Hubble function do not diverge, either. According to [13], again no cosmological finite-time singularity is formed. This remarkable phenomenon can be explained by the influence of the dark fluid viscosity, which compensates for the effect of thermal radiation and the interaction of dark energy with dark matter.

## V. Conclusion

In this paper, we have studied the relevant influence of the thermal effects caused by Hawking radiation on the formation of singularities of the BR type in the FLRW universe. The investigation has carefully considered the combined effect of both the fluid's viscosity and its interaction with dark matter. A related study was carried out for an ideal fluid previously. It was shown in Ref. [40] that accounting for the Hawking thermal radiation near the singularity leads to a qualitative change in the type of singularity. Namely, in a dark universe with finite-time singularities of Types I and III, a transition to a Type II singularity occurred.

Later, in Refs. [41, 42] this problem was considered for a non-ideal fluid, with the result that, taking into account the viscosity property of a dark fluid leads to a transition from a singularity of Type I to one of Type III. Moreover, singularities could be made disappear in some cases.

The novelty of the present paper resides in the fact that coupled viscous dark fluids, with modified log-corrected power-law (EoS) have been introduced, using a new type of generalized entropy function [11]. This modification of the (EoS) is of real interest, in particular, since it is a relevant special case of the seminal, generalized EoS of Ref. [37], which has given rise to very useful results, already.

Viscous cosmology models have recently gained increasing popularity. From a hydrodynamic perspective, this is a completely natural development, since taking into account viscosity coefficients (there are two: shear and volume) physically means departing from the case of an ideal fluid and accounting for deviations from thermal equilibrium up to first order. In a cosmological context, since cosmic fluids are assumed to be spatially isotropic, shear viscosity is usually ignored.

In this article, the coefficient of bulk viscosity is introduced phenomenological. The rationale is that near a singularity, turbulence in a dark fluid occurs. This possibility appears to be the most natural physically, given the expected violent motions near the singularity. Turbulence typically implies the loss of kinetic energy as heat. This loss is often described macroscopically, in terms of bulk viscosity $\zeta$.

In a cosmological context, taking viscosity into account significantly expands the universality of the theory. In the turbulent motions of a cosmic fluid approaching a future singularity, it is quite logical to assume the emergence of turbulent effects [43], therefore, when considering turbulence, it is necessary to take into account the influence of the viscosity of the dark fluid [44].

Different forms of the bulk viscosity term have been here considered. It turns out, that accounting both for the dark fluid viscosity and for its interaction with dark matter, within the framework of the new generalized cosmology, may actually lead to the absence of singularities or, at the very least, to a transition from a hard singularity of Type I to a milder one of Type III. We have thus explicitly proven that thermal radiation highly softens the singularity. The absence of a singularity can be explained by the fact that both the bulk viscosity term in the equation of state and the thermal radiation term in the Friedmann equation are proportional to a power of the Hubble function. Other cosmological models with future singularities and nonlinear logarithmic interactions have been considered in Refs. [45-48].

The question of the mechanism for the origin of dark energy in cosmology is of great interest. Accounting for the viscosity of fluids during the cosmological evolution of the universe is linked to the accelerated expansion of the universe. This is indicated by the fact that the lambda term in the modified equation of state (5) is chosen in the form of bulk viscosity. This emphasizes that the accelerated expansion of the universe is caused by the viscosity of the dark fluid, and vice versa: the property of viscosity is generated by the expansion of the universe [49].

To finish, we should note that the most recent results of astronomical observation are fully consistent with the results for the theoretical models here obtained, which take duly into account the thermal radiation caused by Hawking's radiation [50]. It would be of interest to study thermal effects on singularities in modified gravity, too (see [51, 52], for reviews).

**Acknowledgements**

This work has been partially supported by the program Unidad de Excelencia María de Maeztu CEX2020-001058-M, and by the Catalan Government, AGAUR project 2021-SGR-00171. It has been also supported by the Sophya Kovalevskaya North-West Mathematical Research Center (Immanuel Kant Baltic Federal University) within project no. 075-02-2025-1789 funded by the Ministry of Science and Higher Education of the Russian Federation.